\documentclass[]{spie}
\usepackage[]{graphicx}

\title{Spectrally Pure RF Photonic Source Based on a Resonant Optical Hyper-Parametric Oscillator}

\author{
W. Liang, D. Eliyahu, A. B. Matsko, V. S. Ilchenko, D. Seidel, and L. Maleki
\skiplinehalf OEwaves Inc., 465 N. Halstead Street, Suite 140, Pasadena, California, 91107, USA}

\authorinfo{Send correspondence to A.B.Matsko:
Andrey.Matsko@oewaves.com}

\begin{document}

\maketitle

\begin{abstract}
We demonstrate a free running 10~GHz microresonator-based RF photonic hyper-parametric oscillator characterized with phase noise better than -60~dBc$/$Hz at 10~Hz, -90~dBc$/$Hz at 100~Hz, and -150~dBc$/$Hz at 10~MHz. The device consumes less than 25~mW of optical power. A correlation between the frequency of the continuous wave laser pumping the nonlinear resonator and the generated RF frequency is confirmed. The performance of the device is compared with the performance of a standard optical fiber based coupled opto-electronic oscillator of OEwaves.
\end{abstract}

\keywords{Whispering Gallery Modes, RF Photonic Oscillator, Phase Noise, Spectrally Pure X-band Oscillator, coupled opto-electronic oscillator, opto-electronic oscillator \\ \\
{\small Reprinted from:
Wei Liang, Danny Eliyahu, Andrey B. Matsko, Vladimir S. Ilchenko, David J. Seidel, and Lute Maleki, "Spectrally pure RF photonic source based on a resonant optical hyper-parametric oscillator", in Laser Resonators, Microresonators, and Beam Control XVI, Alexis V. Kudryashov; Alan H. Paxton; Vladimir S. Ilchenko, Editors, Proceedings of SPIE Vol. 8960 (SPIE, Bellingham, WA 2014), 896010; \\ doi: 10.1117$/$12.2044826; $http://dx.doi.org/10.1117/12.2044826$ }}

\section{Introduction}

Stabilized optical frequency combs are promising for generation of spectrally pure radio frequency (RF) signals. The signals are produced by demodulating the optical harmonics of the frequency combs on a fast photodiode. High degree of relative phase locking of the optical harmonics results in generation of spectrally pure RF signals. For example, a mode locked optical frequency comb was used to generate 10~GHz signals with phase noise at the level of -100 dBc/Hz at 1~Hz offset and -179~dBc/Hz at 1~MHz and higher offsets \cite{fortier11np,quinlan13np}. This is 40~dB lower than the phase noise of the best room-temperature electronic oscillators. The output power of the 10~GHz RF source can reach 25~mW when a uni-traveling carrier photodiode is illuminated with picosecond pulses from a repetition-rate-multiplied Ti:sapphire modelocked laser \cite{fortier13ol}.

Hyper-parametric RF photonic oscillators based on high-Q Whispering Gallery Mode (WGM) microresonators are noticeable among optical frequency comb-based RF sources because of their compactness, power efficiency, and low acceleration sensitivity \cite{savchenkov04prl,savchenkov08prl,savchenkov08oe,maleki10ifcs}. The phase noise of a compact WGM resonator based hyper parametric (Kerr comb) RF photonic oscillators is comparable with the noise of existing electronic oscillators, e.g. dielectric resonator oscillators (DRO). For instance, a good DRO, e.g. 8 GHz Hittite HMC-C200, has phase noise of -122 dBc/Hz at 10 kHz and -140 dBc/Hz at 100 kHz offset. In the past we have demonstrated 10 GHz and 35 GHz Kerr comb RF photonic oscillators with phase noise of -115 dBc/Hz at 10 kHz offset and -130 dBc/Hz at 100 kHz offset \cite{savchenkov13ol}.  A unique advantage of the photonic oscillator is in the possibility of increasing of the generation frequency without performance degradation.

Multimode microresonator-based Raman \cite{liang10prl} and Brillouin \cite{geng08ol,grudinin09prl,gross10oe,lee13nc,eggleton13aop} lasers can also be used for RF generation instead of the Kerr frequency combs. The operation of the devices is identical to the optical comb based devices -- the frequency harmonics are demodulated at a fast photodiode to produce a spectrally pure RF signal at the frequency corresponding to the repetition rate of the harmonics. An important difference between the lasers and the hyper-parametric oscillator is that the phase locking of the harmonics have different phase locking mechanisms in the Raman and Brillouin lasers.

The phase locked Raman laser allowed generating 35~GHz RF signals characterized with phase noise of -30~dBc/Hz at 100~Hz offset and -140~dBc/Hz at 10 MHz and higher offsets \cite{liang10prl}. The best phase noise achieved with a microresonator-based Brillouin laser and reported so far (21.7 GHz carrier frequency is set by the resonator host material properties) is -30~dBc/Hz at 100~Hz offset, -90~dBc/Hz at 10~kHz offset, -110~dBc/Hz at 100~kHz offset, and -160~dBc/Hz at 100~MHz and higher frequency offsets \cite{lee13nc}.

An important advantage of the RF photonic oscillators based on the Brillouin lasing is their power efficiency. The output RF power leaving the photodiode exceeds 1~mW without any optical or RF amplification \cite{lee13nc}. The noise floor as well as the output RF power are superior over those previously reported for the microresonator-based RF photonic oscillators.

It is possible to generate RF frequency signals using optical fiber devices with resonators utilized as passive filters. Opto-electronic oscillators (OEOs) and mode locked lasers operate in this way \cite{delfyett06jlt,gee05ptl,quinlan08jlt,quinlan09joa}. The phase noise of a free running 10~GHz harmonically mode-locked semiconductor ring laser with an optical etalon (resonator) is at the level of -90~dBc/Hz at 10~Hz offset, -100~dBc/Hz at 10~kHz offset, and -140~dBc/Hz at 10~MHz and higher frequency offsets \cite{gee05ptl}.

Optical microresonators also can be used as resonant nonlinear elements in fiber lasers \cite{pasquazi13oe,johnson13arxiv}. Such a configuration simplifies locking the system to achieve stable operation. Phase noise of the devices is not studied in detail yet.
\begin{figure}
\centering\includegraphics[width=7.5cm]{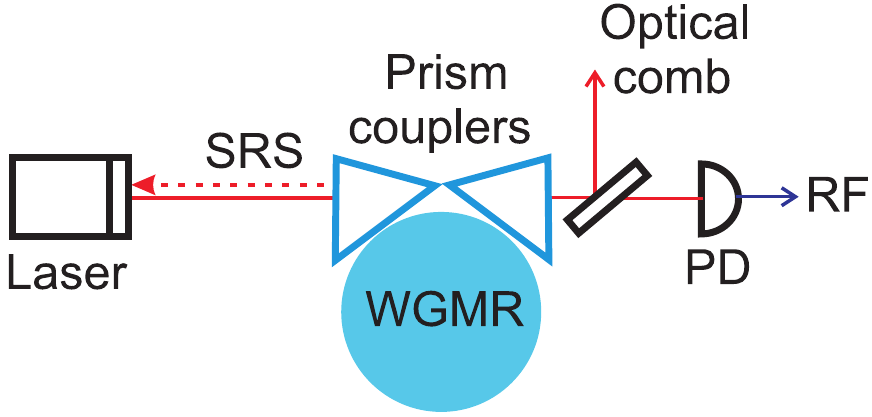}
\caption{\label{fig1KOEO} Schematic of the setup of the RF photonic oscillator based on optical hyper-parametric oscillation in a MgF$_2$ crystalline whispering gallery mode resonator (WGMR). The resonator is pumped with a semiconductor DFB laser using an evanescent prism coupler made of BK7 glass. Since the free running laser linewidth ($\approx 5$~MHz) is much larger as compared with the unloaded bandwidth of the resonator modes, the laser needs to be locked to the resonator mode. The locking is achieved with help of resonant stimulated Rayleigh scattering (SRS) of light in the body of the resonator \cite{liang10ol}. When the pump power exceeds certain threshold, the light exiting the resonator through another BK7 prism coupler is modulated at the frequency approximately equal to the free spectral range of the resonator. The modulation corresponds to generation of the optical frequency comb (Kerr comb). Demodulating the light on a fast photodiode (PD) we generate spectrally pure radio frequency (RF) signals at frequencies corresponding to multiples of the free spectral range of the resonator. }
\end{figure}

In this work we report on the development and fabrication of an RF photonic oscillator based on the Kerr
frequency comb device that demonstrates phase noise order of magnitude lower as compared to the earlier
demonstrations. Namely, our oscillator has phase noise better than -60~dBc/Hz at 10~Hz, -90~dBc/Hz at
100 Hz, and -150~dBc/Hz at 10~MHz. The device consumes less than 25~mW of optical power. The total
power consumption is determined by the thermal stabilization of the platform and is about 2~W. Therefore, the
performance of the oscillator is superior as compared with the performance of existing compact RF photonic
oscillators

We present the measurement results of the Kerr frequency comb oscillator and compare it with performance
of OEwaves’ coupled opto-electronic oscillator in Section II of the paper. The origin of the phase noise limitations
of the comb oscillator is investigated in Section III. In the same section we present a study of correlations between
frequency of the pump laser and frequency of the RF signal. Section VI concludes the paper.

\section{Experiment}
\subsection{RF photonic hyper-parametric oscillator}

We created an RF photonic oscillator based on the optical frequency comb generated in a high-Q MgF$_2$ whispering gallery mode (WGM) resonator. A schematics of the oscillator setup is shown in Fig.~(\ref{fig1KOEO}). A WGM optical resonator is fabricated by mechanical polishing of a cylindrical crystalline preform. The resonator has 7.1~mm in diameter, 0.1~mm in thickness, and 9.9~GHz free spectral range (FSR). The intrinsic (unloaded) bandwidth of the modes is 35~kHz. The loaded (with two BK7 prisms) bandwidth approaches 300~kHz. A semiconductor distributed feedback (DFB) laser is coupled to the resonator through the input prism and locks to a mode through the intrinsic resonant stimulated Rayleigh scattering \cite{liang10ol}. This, so called ”self-injection locking”, results in 30-40~dB improvement of the laser linewidth, which is necessary to couple light from the laser to a high-Q WGM. The locking mechanism is stable even though the nonlinear process modifies the resonant response of the mode rather significantly. The threshold optical power of the frequency comb generation varies from sub mW to a few mW depending on the particular mode selected and the loaded Q. The maximum optical power sent to the resonator is less than 25~mW.
\begin{figure}
\centering\includegraphics[width=14cm]{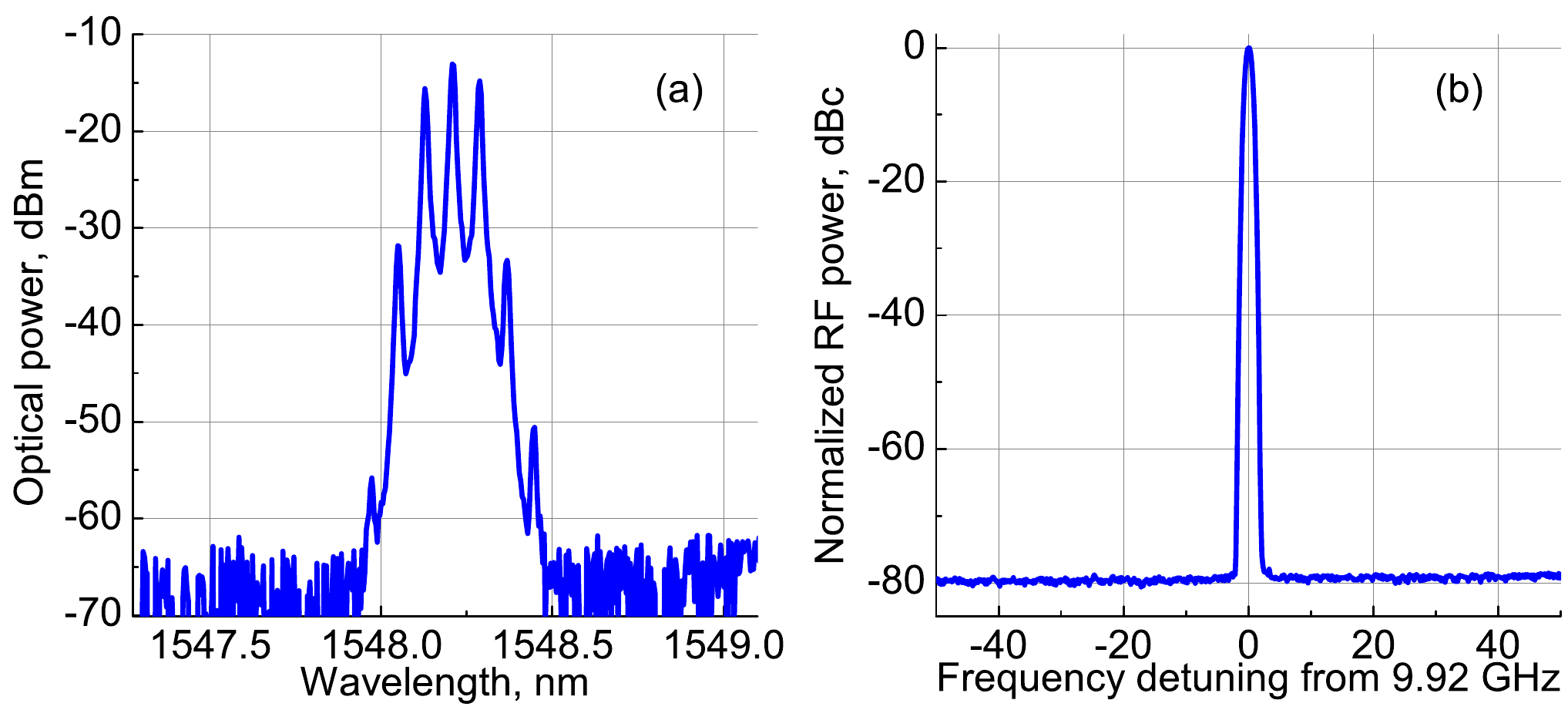}
\caption{\label{fig2KOEO} (a) Spectrum of an optical signal leaving the resonator. (b) Spectrum of the RF signal  leaving the fast photodiode. The signal is measured using an RF spectrum analyzer. The resolution bandwidth of the measurement is 100~MHz. Taking into account the 80~dB contrast of the signal we find that the noise floor of the signal is approximately -160~dBc/Hz at high offset frequencies. }
\end{figure}

Kerr frequency comb starts when the circulating in the mode optical power exceeds a certain threshold. Optical harmonics appear due to modulation instability of the cw background. The linewidth of the harmonics is approximately the same as the linewidth of the pump light. However, their relative frequency fluctuations are strongly correlated. Beating of the optical comb signal leaving the resonator (Fig.~\ref{fig2KOEO}a) on a photodiode results in generation of a spectrally pure radio frequency (RF) signal (Fig.~\ref{fig2KOEO}b).

\begin{figure}
\centering\includegraphics[width=14cm]{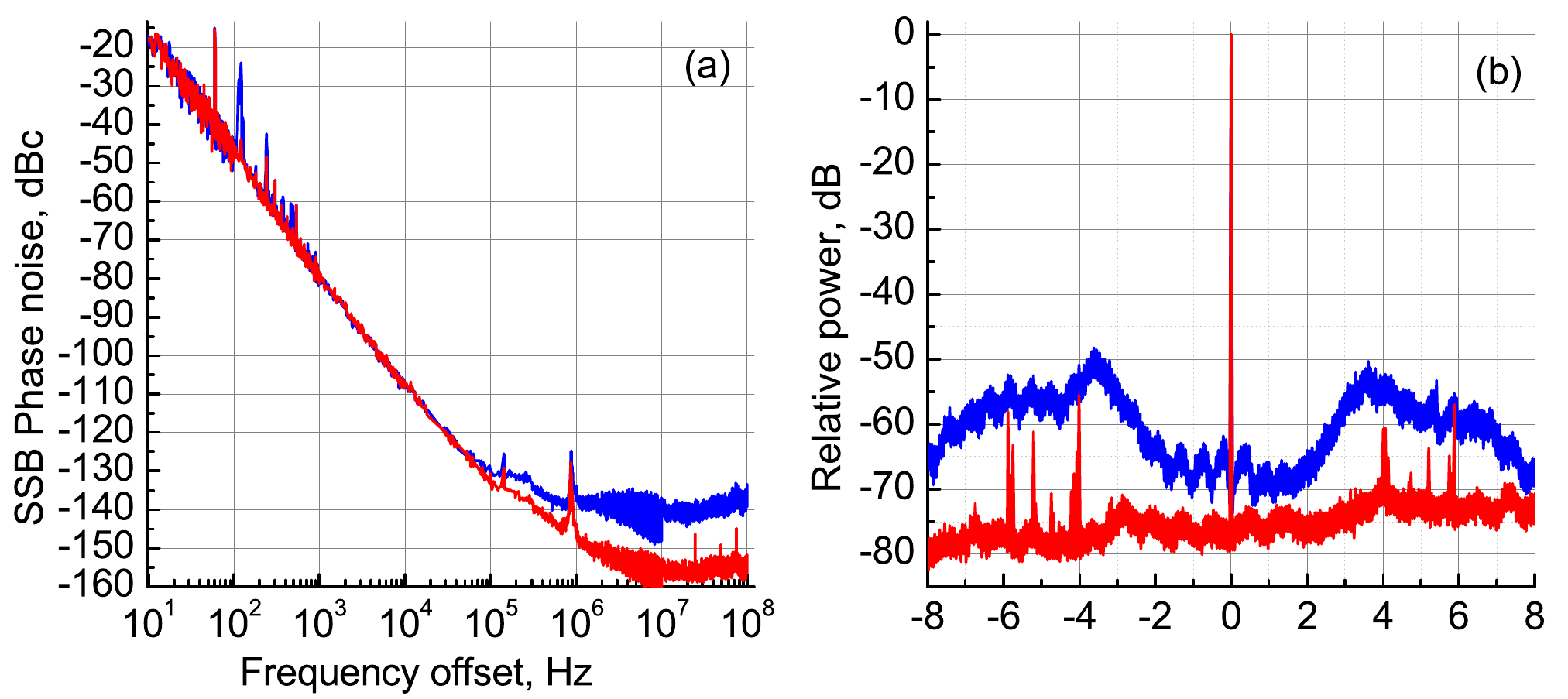}
\caption{\label{fig3KOEO} (a) Single sideband phase noise of Kerr frequency comb oscillator measured in transmission (red) and reflection (blue) ports, respectively. (b) Beat signal between the self-injection locked DFB laser and a Koheras narrow linewidth fiber laser measured with a RF spectrum analyzer. The signal of the DFB laser is retrieved from reflection (blue) and transmission ports (red), respectively. }
\end{figure}

We studied generation of the Kerr comb using one and two prism couplers. Usually the Kerr frequency comb produced has significant energy in the nearest sidebands only (Fig.~\ref{fig2KOEO}a), RF signal generated in this regime already has very good spectral purity.  We noticed that the phase noise floor (blue curve at (Fig.~\ref{fig3KOEO}a) of the RF signal is far from the level dictated by shot noise and relative intensity noise (RIN) of the laser. This noise floor is limited by the residual phase noise of the self injection locked laser. We then generated the RF signal utilizing the comb signal transmitted through the 2nd prism and the phase noise floor was lowered to the almost -160~dBc/Hz limited by shot noise (red curve at Fig.~\ref{fig3KOEO}a). It worth noting that two prism configuration was already used to reduce DC background as well as residual noise in Kerr frequency comb oscillators \cite{matsko13ol,wang13oe}.

We further investigated the phase noise profile of the light produced by self-injection locked DFB laser by mixing it with a narrow linewidth light produced by a fiber laser (Koheras) and observed the RF beat signal on a microwave spectrum analyzer.  We measured the phase noise profiles of the laser beat-note signal (10~GHz frequency) at both the reflection port and the transmission port of the prism pair. The results are plotted in Fig.~\ref{fig3KOEO}b. The phase noise of the signal that involves in-transmission light is further lowed by the filtering of the resonator, which enables us to achieve shot noise limited phase noise of the RF signal (see also Fig.~\ref{fig2KOEO}b).

\begin{figure}
\centering\includegraphics[width=14cm]{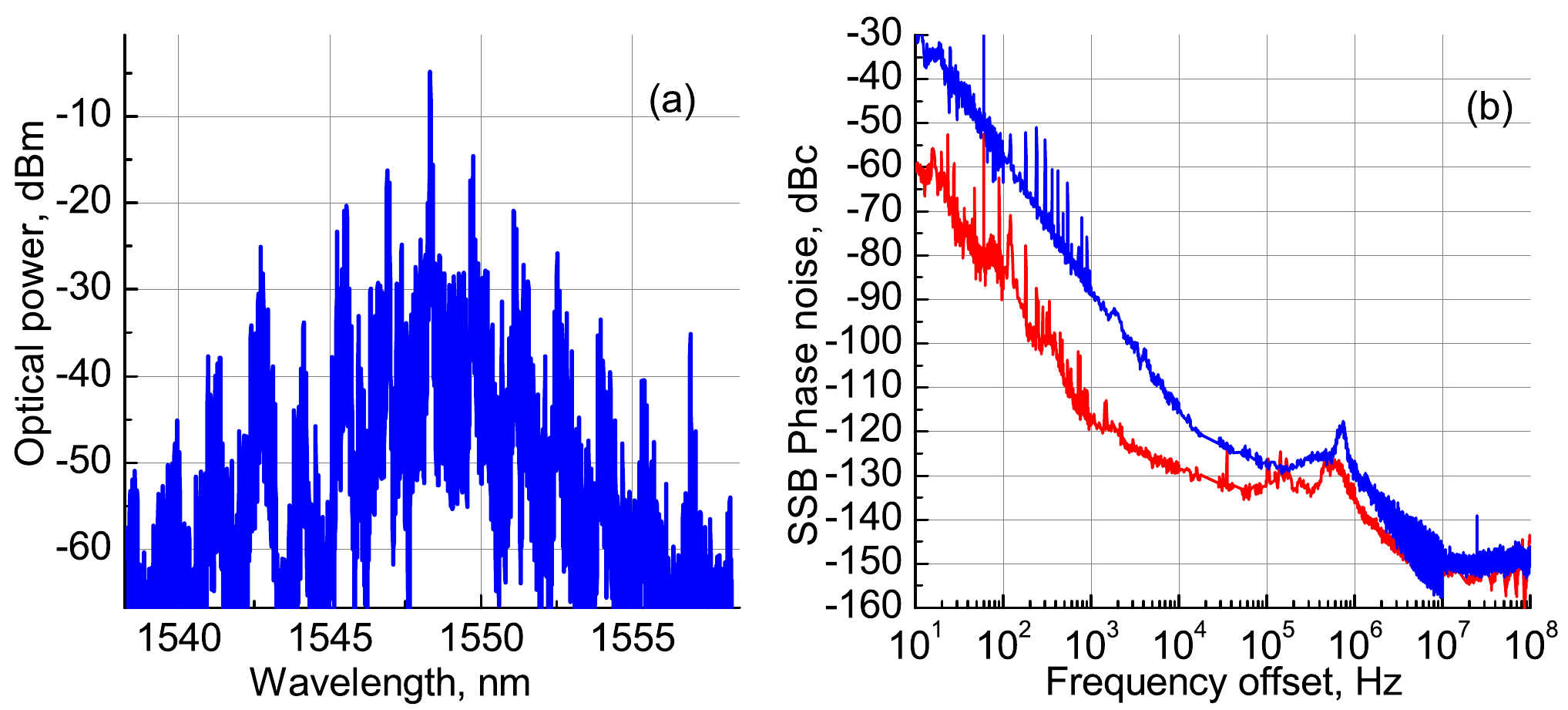}
\caption{\label{fig4KOEO} (a) Optical spectrum of a Kerr frequency comb spanning a wide range of wavelength. (b) Phase noise of the RF signal generated by the wide Kerr frequency comb. The two curves were obtained for different experimental parameters.}
\end{figure}

By selecting a different mode family of the multimode WGM resonator and changing the pump laser power, we were able to generate a Kerr frequency comb of properties different from the one aforementioned. In essence, the optical spectrum of the comb spans a much wider frequency range (Fig.~\ref{fig4KOEO}a) with RF beat signals of much lower close-in phase noise (Fig.~\ref{fig4KOEO}b). A slight change of the experimental parameters can result in rather significant modification of the performance of the system (compare red and blue curves in Fig.~\ref{fig4KOEO}b). By optimizing the experimental condition, we were able to achieve phase noise level of -60~dBc/Hz at 10~Hz, -90~dBc/Hz at 100~Hz, and -150~dBc/Hz at 10~MHz.

\subsection{Coupled opto-electronic oscillator (COEO)}

\begin{figure}
\centering\includegraphics[width=9cm]{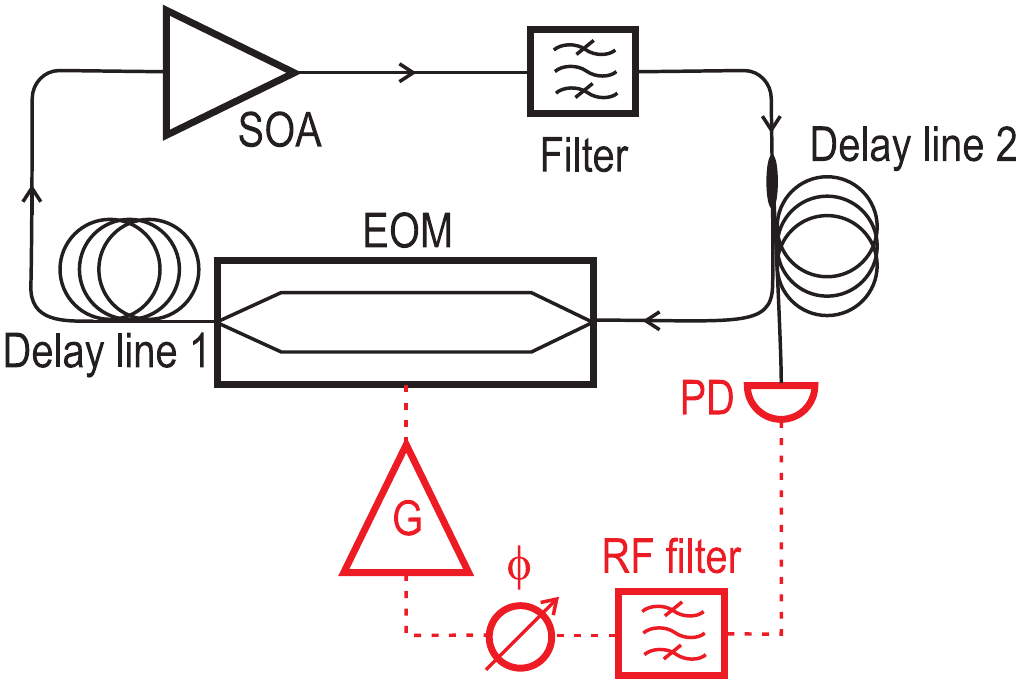}
\caption{\label{fig1coeo} Schematic diagram of the coupled opto-electronic oscillator (COEO). Optical and RF loops are shown by solid black and red dashed lines, respectively. Both loops include amplifiers and filters. The optical loop contains a semiconductor optical amplifier(SOA). The RF loop also includes an RF amplifier. A part of the light circulating in the optical loop is demodulated on a fast photodiode (PD) generating an RF signal with carrier frequency equivalent to the optical pulse repetition rate. The amplified and filtered RF signal is fed into an electro-optical modulator (EOM). The phase of the RF signal is tuned via the phase rotator $\phi$. The oscillator contains two optical delay lines, one in the optical and the other in the RF feedback loops. The dispersions of the delay lines have to be optimized independently. The nonlinearity of the delay line in the optical loop should be taken into account to understand mode locking in the device, while the nonlinearity of the second delay line is generally not important because of the comparably low circulating optical power.}
\end{figure}

It is interesting to compare the phase noise of the Kerr frequency comb oscillator to that of a commercial COEO units (OEwaves Compact Opto-Electronic Oscillator) (see \cite{matsko09josab,matsko13josab} for details) as the COEOs are among the best existing commercially available RF oscillators. The COEO units have architecture as shown in Fig.~(\ref{fig1coeo}). A semiconductor optical amplifier is used as the amplifying medium. The amplifier is characterized with unsaturated gain of 25~dB, saturation energy of 1~pJ, and gain recovery time of 0.1~ns. The average dispersion in the fiber loop is around $-8$~ps$/$nm$\;$km with a nonlinearity coefficient of $5$~W$^{-1}$km$^{-1}$. An electro-absorption amplitude modulator as well as an optical filter are utilized  in these COEO units. The modulator is characterized with $V_{\pi}$ of 5~V, with an insertion loss of $-5$~dB. A portion of the light propagating in the optical cavity is coupled out and downconverted to 10 GHz signal. This signal is filtered by a narrow band RF filter, amplified and then sent back to drive the electro-optical modulator.

\begin{figure}
\centering\includegraphics[width=15cm]{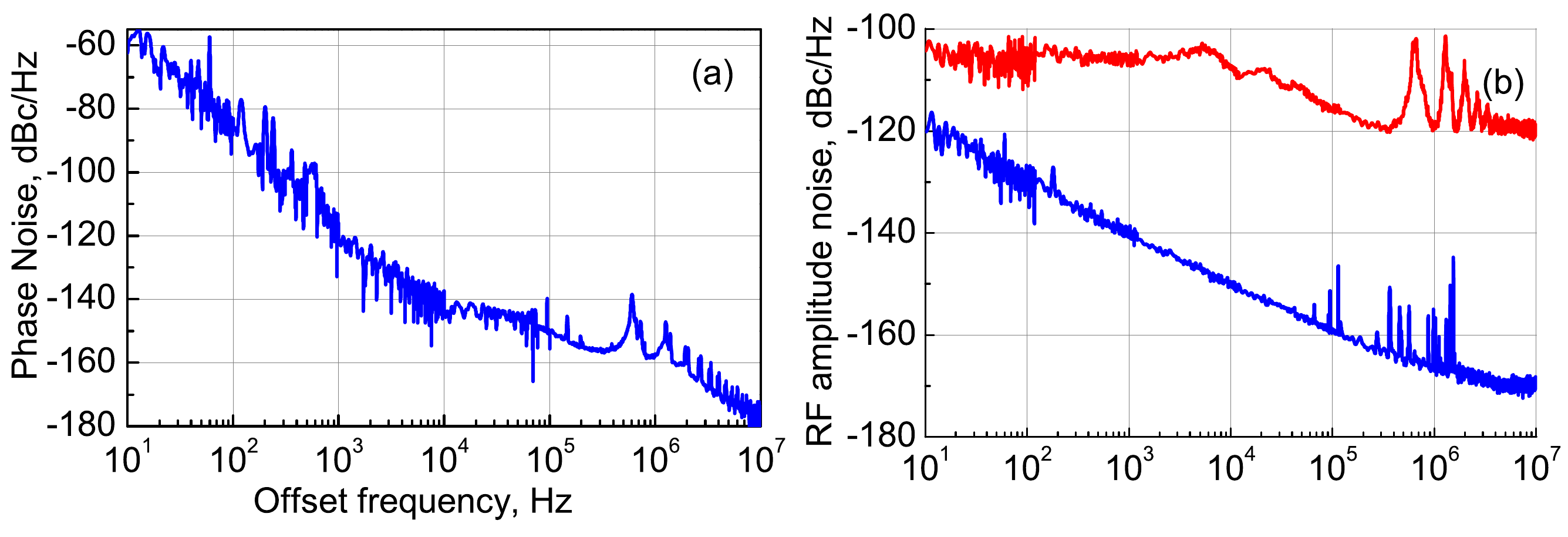}
\caption{\label{fig2coeo} (a) Typical phase noise of experimentally demonstrated COEO.  The peaks below 100~kHz in the experimental spectrum are artifacts of the measurement system. (b) Amplitude noise measured for COEO operating in mode locked (blue line) and not mode locked (red line) regimes. The noise drops significantly when the mode locked regime is achieved.}
\end{figure}

An example of experimentally observed phase noise is shown in Fig.~(\ref{fig2coeo}a). Close-in noise behavior can be explained by flicker noise of the amplifiers in the loop, as well as the thermal stabilization efficiency of the fiber. Also amplified spontaneous emission noise of the optical amplifier has to be taken into account to explain the noise floor.

To verify if the excessive phase noise comes from the conversion of the amplitude to phase noise within the COEO loop we measured the amplitude noise of the RF signal generated by the device (see Fig.~\ref{fig2coeo}b). While the amplitude noise of the unlocked COEO is high, the mode locking reduces the noise significantly, well below the phase noise. Therefore, it is unlikely that the amplitude noise can explain the excessive phase noise.

\begin{figure}
\centering\includegraphics[width=15cm]{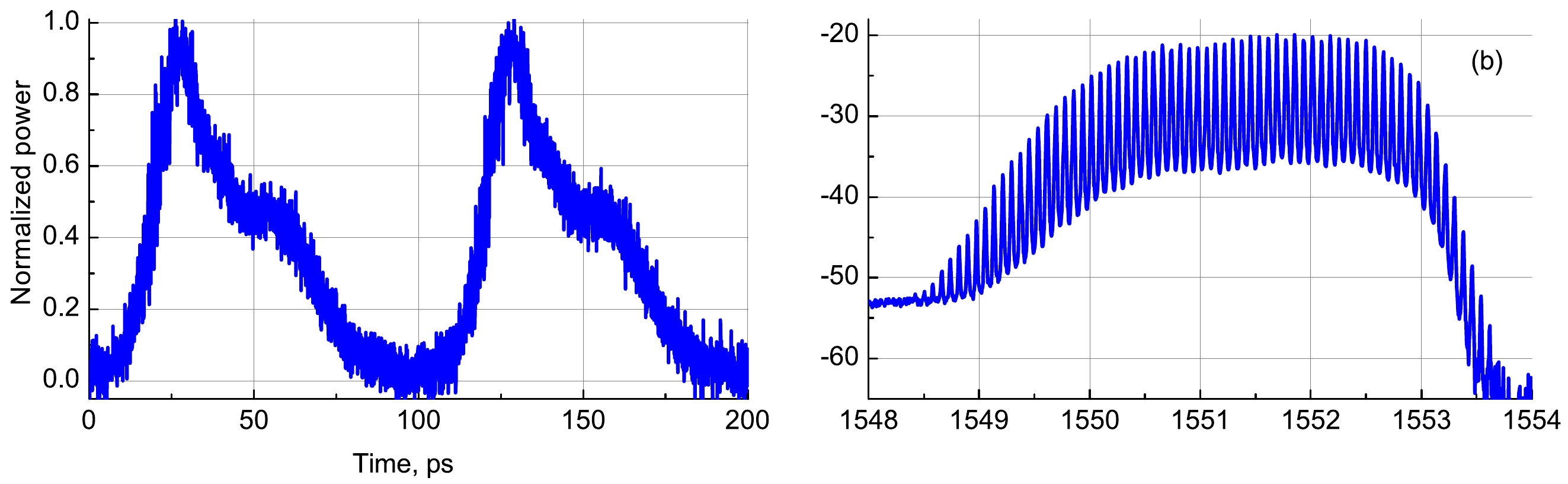}
\caption{\label{fig3coeo} Typical optical pulses (a) and associated optical spectrum (b) generated by a COEO. The pulse shape was measured using an Agilent wide bandwidth oscilloscope.}
\end{figure}

The optical pulses generated in the optical loop have an irregular shape (Fig.~\ref{fig3coeo}a). The shape distortion comes from the slow carrier recombination time in the semiconductor optical amplifier. We expect that usage of a fiber amplifier instead of the semiconductor amplifier would improve performance of the oscillator and will result in reduction of the pulse duration and pulse shape improvement.
The optical spectrum of the COEO is a nicely shaped nearly flat frequency comb (Fig.~\ref{fig3coeo}b). While generated pulses are comparably long, resulting in an imperfect the mode locking regime, the phase locking is very good, which is supported by the phase noise of the RF signal generated by the comb on a fast photodiode.

\section{Discussion}

Comparing data shown at Fig.~(\ref{fig3KOEO}) and at Fig.~(\ref{fig2coeo}a) we find that the close-in noise of the Kerr frequency comb-based  device is already at the same level as the close-in noise of the fiber-based COEO. This is an impressive result as the size of the microresonator-based comb oscillator package is two orders of magnitude smaller as compared with the size of the fiber-based device. The question is what limits the comb oscillator performance and if it is possible to create a comb oscillator with performance better than the COEO one.

\begin{figure}
\centering\includegraphics[width=14cm]{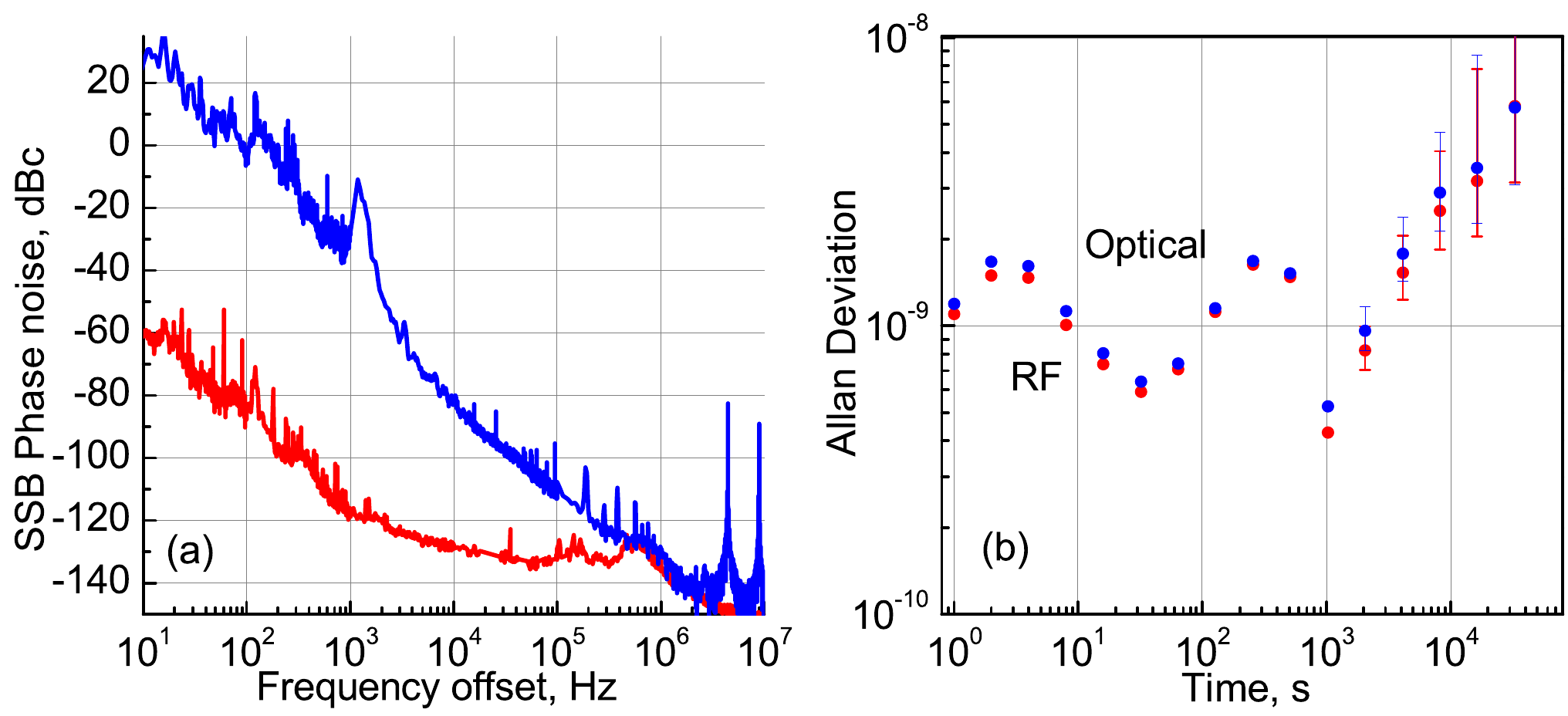}
\caption{\label{fig1discussion} (a)  Comparison between the phase noises of an injection locked DFB laser and the generated Kerr frequency comb RF signal. The spurs of the laser phase noise at offset frequency smaller than 1~kHz are from the acoustic noise picked up by the fiber interferometer used to measure the phase noise. The difference between the phase noise values is about 86~dB at close-in frequency, which corresponds to $20{\rm log}(\nu_0/\nu_{RF})$ (b) Allan deviation evaluated for the carrier frequency of the cw light pumping the nonlinear WGM resonator, $\nu_0$, and for the repetition frequency of the optical comb oscillator, $\nu_{RF}$. Overlap of the Allan deviation values means that the ratio of the corresponding phase noise values of the signals is $\nu_0^2/\nu_{RF}^2$. }
\end{figure}

To answer this question we have performed two different experiments. In the first one we used an OEwaves laser frequency noise measurement system to detect the phase noise of the DFB laser while it is self-injection locked to the WGM resonator and generating a wide Kerr frequency comb. In (Fig.~\ref{fig1discussion}a) we compare the laser phase noise with the best phase noise of the Kerr frequency comb oscillator. The phase noise of the laser was measured with a frequency discriminator made of fiber Mach-Zender interferometer (MZI). The peaks of the phase noise curve for offset frequency smaller than 1~kHz are mostly coming from acoustic noise picked by the fiber MZI. The intrinsic phase noise of the laser is outlined by the dashed line. The difference between the two curves at close-in frequency is about 86~dB, which is exactly 20log of the ratio between the optical frequency ($\nu_0 =200$~THz) and the FSR frequency ($\nu_{RF}=9.9$~GHz). Hence, we can conclude that at the optimal operation condition, the phase noise of comb oscillator is limited by the phase noise of the injection locked laser divided by the frequency ratio squared, $\nu_0^2/\nu_{RF}^2$. The noise originates from the thermal noise of the WGM resonator.

In the second experiment we have measured a correlation between frequencies of the pump light and the RF signal \cite{savchenkov13ol}. Since the drift of the laser frequency is primarily given by the drift of the resonator mode the laser is locked to, this kind of a measurement shows if the comb repetition frequency drift depends on the overall drift of the optical mode frequency. We can expect such a behavior from the phase noise measurement discussed above.

To measure the optical frequency we utilized the heterodyne technique and used an optical local oscillator (LO) made by locking a semiconductor laser to an HCN optical transition. The measured 1~s stability of the optical LO was on the order of $10^{-10}$, so we had to reduce the stability of our RF photonic oscillator to perform the measurement. We switched off the direct temperature stabilization of the resonator (only the temperature of the platform was stabilized) and observed simultaneous drift of the pump laser frequency and comb repetition rate (RF frequency). RF and optical frequencies were simultaneously recorded for one day and then compared.

The result of this measurement is shown in (Fig.~\ref{fig1discussion}b). The optical and RF frequencies drifted together and the ratio of their drifts was constant and equal to the ratio of the FSR and optical frequencies. This measurement also shows that the close-in phase noise of both the laser and the resonator phase is given by the thermal drift of the resonator mode. The degree of correlation has to be studied in more details. However, even at this point, we can conclude that the close-in phase noise of the RF photonic oscillator is strongly influenced by the thermal drifts of the WGM resonator. Stabilization of the resonator temperature should reduce the noise.

\section{Conclusion}
We have demonstrated a WGM-based RF photonic Kerr frequency comb oscillator with low phase noise value approaching phase noise value of a fiber-based coupled opto-electronic oscillator. We found that the close-in phase noise of the comb oscillator depends on the frequency drifts of the nonlinear whispering gallery mode resonator. Further thermal and mechanical stabilization of the resonator will result in improvement of the RF phase noise.

\section*{Acknowledgments}
The authors acknowledge support from Defense Sciences Office of Defense Advanced Research Projects Agency under contract No. W911QX-12-C-0067 as well as support from Air Force Office of Scientific Research under contract No. FA9550-12-C-0068.




\begin{thebibliography}{99}
\frenchspacing

\bibitem{fortier11np} T. M. Fortier M. S. Kirchner, F. Quinlan, J. Taylor, J. C. Bergquist, T. Rosenband, N. Lemke, A. Ludlow, Y. Jiang, C. W. Oates, and S. A. Diddams, "Generation of ultrastable microwaves via optical frequency division," Nature Photonics {\bf 5}, 425-429 (2011).

\bibitem{quinlan13np} F. Quinlan, T. M. Fortier, H. Jiang, A. Hati, C. Nelson, Y. Fu, J. C. Campbell, and S. A. Diddams, "Exploiting shot noise correlations in the photodetection of ultrashort optical pulse trains," Nature Photonics {\bf 7}, 290–-293 (2013).

\bibitem{fortier13ol} T. M. Fortier, F. Quinlan, A. Hati, C. Nelson, J. A. Taylor, Y. Fu, J. Campbell, and S. A. Diddams, "Photonic microwave generation with high-power photodiodes," Opt. Lett. {\bf 38}, 1712-1714 (2013).

\bibitem{savchenkov04prl} A. A. Savchenkov, A. B. Matsko, D. Strekalov, M. Mohageg, V. S. Ilchenko, and L. Maleki, "Low threshold optical oscillations in a whispering gallery mode CaF$_2$ resonator," Phys. Rev. Lett. {\bf 93}, 243905 (2004).

\bibitem{savchenkov08prl} A. A. Savchenkov, A. B. Matsko, V. S. Ilchenko, I. Solomatine, D. Seidel, and L. Maleki, "Tunable optical frequency comb with a crystalline whispering gallery mode resonator," Phys. Rev. Lett. {\bf 101}, 093902 (2008).

\bibitem{savchenkov08oe} A. A. Savchenkov, E. Rubiola, A. B. Matsko, V. S. Ilchenko, and L. Maleki, ”Phase noise of whispering gallery photonic hyper-parametric microwave oscillators,” Opt. Express {\bf 16}, 4130-4144 (2008).

\bibitem{maleki10ifcs} L. Maleki, V. S. Ilchenko, A. A. Savchenkov, W. Liang, D. Seidel, and A. B. Matsko, "High performance, miniature hyper-parametric microwave photonic oscillator," Proc. 2010 IEEE International Frequency Control Symposium, pp. 558 -- 563 (2010).

\bibitem{savchenkov13ol} A. A. Savchenkov, D. Eliyahu, W. Liang, V. S. Ilchenko, J. Byrd, A. B. Matsko, D. Seidel, and L. Maleki, "Stabilization of a Kerr frequency comb oscillator," Opt. Lett. {\bf 38}, 2636-2639 (2013).

\bibitem{liang10prl} W. Liang, V. S. Ilchenko, A. A. Savchenkov, A. B. Matsko, D. Seidel, and L. Maleki. "Passively mode-locked Raman laser," Physical Review Letters {\bf 105}, art. no. 143903 (2010).

\bibitem{geng08ol} J. H. Geng, S. Staines, and S. B. Jiang, "Dual-frequency Brillouin fiber laser for optical generation of tunable low-noise radio frequency/microwave frequency," Opt. Lett. {\bf 33}, 16–18 (2008).

\bibitem{grudinin09prl} I. S. Grudinin, A. B. Matsko, and L. Maleki, "Brillouin Lasing with a CaF2 Whispering Gallery Mode Resonator," Phys. Rev. Lett. {\bf 102}, 043902 (2009).

\bibitem{gross10oe} M. C. Gross, P. T. Callahan, T. R. Clark, D. Novak, R. B. Waterhouse, and M. L. Dennis, "Tunable millimeter-wave frequency synthesis up to 100~GHz by dual-wavelength Brillouin fiber laser," Opt. Express {\bf 18}, 13321-13330 (2010).

\bibitem{lee13nc} J. Li, H. Lee, and K. J. Vahala. "Microwave synthesizer using an on-chip Brillouin oscillator," Nature Communications {\bf 4}, art. no. 2097 (2013).

\bibitem{eggleton13aop} B. J. Eggleton, C. G. Poulton, and R. Pant "Inducing and harnessing stimulated Brillouin scattering in photonic integrated circuits," Advances in Optics and Photonics {\bf 5}, 536-587 (2013).

\bibitem{delfyett06jlt} P. J. Delfyett, S. Gee, M. T. Choi, H. Izadpanah, W. Lee, S. Ozharar, F. Quinlan, and T. Yimaz, "Optical frequency combs from semiconductor lasers and applications in ultrawideband signal processing and communications," J. Lightwave Technol. {\bf 7}, 2701-2719 (2006).

\bibitem{gee05ptl} S. Gee, F. Quinlan, S. Ozharar, and P. J. Delfyett, "Simultaneous optical comb frequency stabilization and super-mode noise suppression of harmonically mode-locked semiconductor ring laser using an intracavity etalon," IEEE Photon. Technol. Lett. {\bf 17}, 199-201 (2005).

\bibitem{quinlan08jlt} F. Quinlan, C. Williams, S. Ozharar, S. Gee, and P. J. Delfyett, "Self-Stabilization of the optical frequencies and the pulse repetition rate in a coupled optoelectronic oscillator," J. Lightwave Technol. {\bf 26}, 2571-2577 (2008).

\bibitem{quinlan09joa} F. Quinlan, S. Ozharar, S. Gee, and P. J. Delfyett, "Harmonically mode-locked semiconductor-based lasers as high repetition rate ultralow noise pulse train and optical frequency comb sources," J. Opt. A: Pure Appl. Opt. {\bf 11}, 103001 (2009).

\bibitem{pasquazi13oe} A. Pasquazi, L. Caspani, M. Peccianti, M. Clerici, M. Ferrera, L. Razzari, D. Duchesne, B. E. Little, S. T. Chu, D. J. Moss, and R. Morandotti, "Self-locked optical parametric oscillation in a CMOS compatible microring resonator: a route to robust optical frequency comb generation on a chip," Opt. Express {\bf 21}, 13333-13341 (2013).

\bibitem{johnson13arxiv} A. R. Johnson, Y. Okawachi, M. R. E. Lamont, J. S. Levy, M. Lipson, and A. L. Gaeta, "Microresonator-based comb generation without an external laser source," Opt. Express {\bf 22}, 1394-1401 (2014).

\bibitem{liang10ol} W. Liang, V. S. Ilchenko, A. A. Savchenkov, A. B. Matsko, D. Seidel, and L. Maleki, "Whispering-gallery-mode-resonator-based ultranarrow linewidth external-cavity semiconductor laser," Opt. Lett. {\bf 35}, 2822-2824 (2010).

\bibitem{matsko13ol} A. B. Matsko, W. Liang, A. A. Savchenkov, and L .Maleki, "Chaotic dynamics of frequency combs generated with continuously pumped nonlinear microresonators," Optics Letters {\bf 38}, 525-527 (2013).

\bibitem{wang13oe} P.-H. Wang, Y. Xuan, L. Fan, L. T. Varghese, J. Wang, Y. Liu, X. Xue, D. E. Leaird, M. Qi, and A. M. Weiner, "Drop-port study of microresonator frequency combs: power transfer, spectra and time-domain characterization," Opt. Express {\bf 21}, 22441-22452 (2013).

\bibitem{matsko09josab} A.B. Matsko, D. Eliyahu, P. Koonath, D. Seidel, and L. Maleki, "Theory of coupled optoelectronic microwave oscillator I: expectation values," J. Opt. Soc. Am. B {\bf 26}, 1023-1031 (2009).

\bibitem{matsko13josab} A. B. Matsko, D. Eliyahu, and L. Maleki, "Theory of coupled optoelectronic microwave oscillator II: phase noise," J. Opt. Soc. Am. B {\bf 30}, 3316-3323 (2013).




\end{thebibliography}
\end{document}